\documentclass[useAMS]{mn2e}

\usepackage[ansinew]{inputenc}
\usepackage{graphicx} 

\title[Nuclear and Extended Spectra of NGC 1068]{Nuclear and Extended Spectra of NGC 1068 - II: Near-Infrared Stellar Population Synthesis.}
\author[Martins et al.]{Lucimara P. Martins$^{1}$\thanks{E-mail:
lucimara.martins@cruzeirodosul.edu.br},Rog\'erio Riffel$^{2}$,  Alberto Rodr\'{\i}guez-Ardila$^3$,\newauthor  Ruth Gruenwald$^{4}$, and Ronaldo de Souza$^{4}$\\
$^{1}$NAT - Universidade Cruzeiro do Sul, Rua Galvao Bueno, 868, S\~ao Paulo, SP, Brazil\\
$^{2}$Universidade Federal do Rio Grande do Sul - IF,
Departamento de Astronomia, CP 15051, 91501-970, Porto Alegre, RS
Brasil. \\
$^{3}$Laborat\'orio Nacional de Astrof\'isica/MCT, Rua dos Estados Unidos 154, CEP~37501-064. Itajub\'a, MG, Brazil\\
$^{4}$Instituto Astron\^omico e Geof\'isico - USP, Rua do Mat\~ao, 1226, S\~ao Paulo, SP}

\begin{document}

\date{Accepted ? December ? Received ? December ?; in original form ? October ?}

\pagerange{\pageref{firstpage}--\pageref{lastpage}} \pubyear{2009}

\maketitle

\label{firstpage}

\begin{abstract}

We performed stellar population synthesis on the nuclear and extended 
regions of NGC~1068 by means of near-infrared spectroscopy to disentangle 
their spectral energy distribution components. 
This is the first time that such a technique is applied
to the whole 0.8$-$2.4~$\mu$m wavelength interval in this galaxy. NGC~1068 is one of the nearest
and probably the most studied Seyfert 2 galaxy, becoming an excellent
laboratory to study the interaction between black holes, the
jets that they can produce and the medium in which they propagate.
Our main result is that traces of young stellar population are found at $\sim$100~pc
south of the nucleus. 
The contribution of a power-law
continuum in the centre is about 25$\%$,
which is expected if the light is scattered from a Seyfert 1 nucleus. We find
peaks in the contribution of the featureless continuum 
about 100 - 150 pc from the nucleus on both sides.
They might be associated with regions where the
jet encounters dense clouds. Further support to this scenario
is given by the peaks of hot dust distribution found around these same regions and the
H$_2$ emission line profile,  leading us to propose that the peaks 
might be associate to regions where stars are being formed. Hot dust 
also has an important contribution to the nuclear region, 
reinforcing the idea of the presence of a dense, circumnuclear
torus in this galaxy. Cold dust appears mostly in the south
direction, which supports the view that the southwest emission
is behind the plane of the galaxy and is extinguished very likely by
dust in the plane. Intermediate age stellar population contributes
significantly to the continuum, specially in the inner
200~pc.

\end{abstract}


\section{Introduction}

The coexistence of black holes and starburst clusters is known to exist
in many galaxies, and there are many evidences that suggest a connection
between these phenomena. Many studies point out that both the active
nucleus and starbursts might be related to gas inflow, probably triggered by an axis-asymmetry 
perturbation like bars, mergers or tidal interactions (Shlosman, Frank \& Begelman
1989, Shlosman, Begelman \& Frank 1990, Maiolino et al. 1997, Knapen, Shlosman \& Peletier
2000, Fathi et al. 2006, Riffel et al. 2008).
In addition, one of the most intriguing research areas in contemporary 
extragalactic astrophysics involves the study of the interplay between
nuclear black holes, the jets which they can produce and the interstellar/intergalactic
medium (ISM) in which they propagate. These jets can have a considerable
impact on this medium. One aspect of jet-ISM interaction
is that it can trigger star formation.
Such jet-induced star formation is considered a possible mechanism to explain
the UV continuum emission observed in the host galaxies of distant radio sources and the
``alignment effect'' between the radio emission and this continuum (Rees
1989). 
Although this effect might play a very important role in high-z radio
galaxies, detecting and studying the jet-ISM interaction in them is very challenging. 
Because of the observational problems, it
is important to find nearby examples of this kind of interaction where a
detailed study can be carried out.

NGC 1068 is an ideal object in this case. It is one of the nearest and probably the most
intensely studied Seyfert 2 galaxy. Observations in all wavelength bands
from radio to hard X-rays have formed a
uniquely detailed picture of this source. NGC 1068 hosts a
prominent narrow-line region (NLR) that is approximately
co-spatial with a linear radio source with two lobes (Wilson
\& Ulvestad 1983). Star formation activity
coexistent with the active galactic nucleus (AGN) was detected on both
larger (e.g., Telesco \& Decher 1988) and smaller scales (Macchetto
et al. 1994; Thatte et al. 1997). However, the link between
all the processes is still under debate.

The Near Infrared Region (NIR) is particularly interesting to help
unveiling this link because it is accessible to
ground-based telescopes and, at the same time, able to probe highly obscured sources. 
However, tracking the star formation in the NIR is not simple (Origlia \& Oliva
2000) although recent studies exploring this
region have already shown its strong potential at detecting 
intermediate-age stellar population not easily tracked 
in the optical without
ambiguity (e.g. Riffel et al. 2009, Davies et al. 2007).
At near-IR wavelengths stellar photospheres usually remain the dominant 
sources of light, and galaxy spectra are shaped by red supergiants (RSG) 
shortly after starbursts, and then by giants of the first and of the 
asymptotic giant branches (AGB). 

AGB stars are rare members of stellar populations. However,
they are among the most luminous cool stars and can therefore
be detected sometimes even individually in galaxies.
The TP-AGB stars leave a unique fingerprint on the integrated spectra,
like the 1.1 $\micron$ CN band (Maraston 2005, Riffel et al. 2007, 2009). 
Hence when detected, they can help to determine the age of the 
stellar population through the integrated light.
The contribution of this stellar phase in stellar population
models has been recently included in both the energetics and the spectral
features (Maraston 2005). In particular these models employ 
empirical spectra of oxygen-rich and carbon stars (Lan\c con 
\& Wood 2000), which are able to foresee characteristic NIR absorption
features.

With this in mind, we present here for the first time in the literature
a detailed fitting of the continuum emission components in the 0.8$-$2.4 $\micron$ 
interval of NGC~1068 across the central 15" ($\sim$ 1100 pc) of this source.
The main purpose is to determine the fraction with which the different components
contributes to the observed integrated light and how they 
are related to each other. The paper
is structured as follows: in \S 2 we describe the observations. In
\S 3 we describe the fitting method and in \S 4 the results are presented
and discussed. Final remarks are given in \S 6.

\section{The Observations}

The spectra were obtained at the NASA 3m Infrared
Telescope Facility (IRTF) in October 30, 2007. The
SpeX spectrograph (Rayner et al., 2003) was used in the short
cross-dispersed mode (SXD, 0.8- 2.4 $\mu$m).  The employed detector
consisted
of a 1024x1024 ALADDIN 3 InSb array with a spatial scale
of 0.15"/ pixel. A 0.8"x 15" slit oriented in the north-south direction
was used, providing a spectral
resolution of 360 km/s. For more details about the instrumental
configuration see Martins et al. (2010, hereafter Paper~I).

Figure~1 shows the position of the slit superimposed
on the galaxy contours obtained from Galliano et al. 
(2003). The gray contours show the 6 cm emission
(Gallimore et al. 1996) and the red dotted contours
show the 20 $\micron$ image (Alloin et al. 2000).
For NGC1068, 17 extractions were made along the spatial
direction: one centred at the peak of light distribution (nuc)
and eight
more at each side of it (apertures 01 to 08 in the south direction
and 09 to 16 in the north direction). Figures 5 to 7 of Paper~I
show the individual extractions along
the spatial direction as well as the most important emission and
absoprtion features relevant to this work.

\begin{figure} 
\includegraphics [width=90mm]{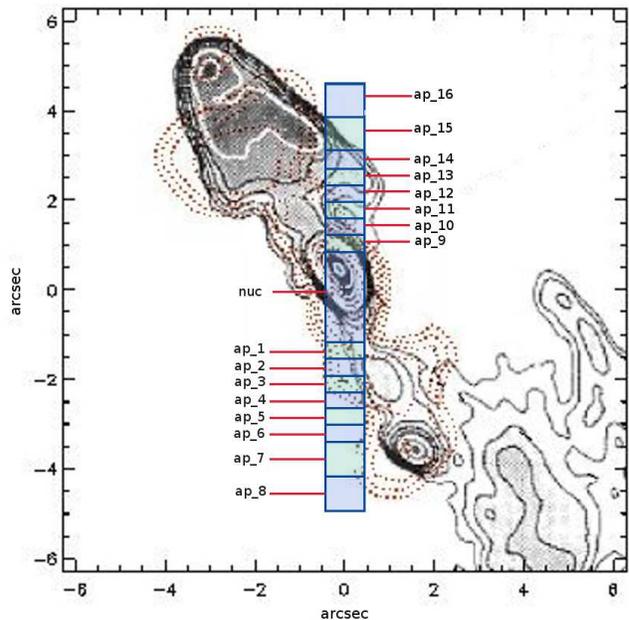}
\caption{NGC 1068 contours obtained from Galliano et al. (2003) with the 
position of the IRTF slit superimposed. The gray contours show the 6 cm emission 
(Gallimore et al. 1996) and the red dotted contours show the 20 $\micron$ image
(Alloin et al. 2000).}
\end{figure}

\section{Spectral Synthesis}

Our main goal  is to study the NIR spectral 
energy distribution components of NGC~1068
and their variations across the central 15".
For this purpose we fit the underlying continuum between 0.8 and 2.4 $\micron$ 
applying the same method described in Riffel et al. 
(2009).

The spectral synthesis is done using the code STARLIGHT
(Cid Fernandes et al. 2004, 2005a, Mateus et al. 2006, Asari et al. 2007,
Cid Fernandes et al. 2009). STARLIGHT mixes computational techniques
originally developed for semi empirical population synthesis with 
ingredients from evolutionary synthesis models.
Basically the code fits an observed spectrum O$_\lambda$ with a combination,
in different proportions, of a number of simple stellar populations (SSPs). 
Due to the fact that the Maraston (2005) models include the effect of the TP-AGB
phase, crucial to model the stellar population in the NIR, we use these SSPs models
as the base set for STARLIGHT. The SSPs used in this work cover
12 ages, t = 0.01, 0.03, 0.05, 0.1, 0.2, 0.5, 0.7, 1, 2, 5, 9
and 13 Gyr, and 4 metalicities, Z = 0.02, 0.5, 1 and 2 z$_{\sun}$,
summing up 48 SSPs.

In addition, in trying to describe the continuum of an AGN, the central
engine cannot be ignored, even outside the nucleus. The stellar population alone
cannot account for the bump in the K band, clearly seen in most of the apertures. Besides that, 
studies have shown the presence of scattered light in NGC~1068 (e.g. Miller, 
Goodrich \& Mathews 1991, Inglis et al. 1995, Simpson et al. 2002).
Usually this component is represented by
a featureless continuum (FC, eg Koski 1978) of power law form that follows
the expression F$_\nu$ $\propto$ $\nu^{-1.5}$ (We have also
tested other indexes for the power law - see Section 4). Therefore, this component 
was also added to the base of elements.

Besides that, the thermal emission from hot dust plays 
an important role in the continuum emission studied in this paper (see Fig. 3 of Riffel
et al. 2009).  Previous studies 
report a mininum in the continuum emission around 1.2 $\micron$ probably associated with
the end of the optical continuum related to the central engine and the onset
of the emission due to reprocessed nuclear radiation by dust (Barvanis 1987,
Thompson 1995, Rudy et al. 2000, Rodr\'iguez-Ardila \& Viegas 2003, Rodr\'iguez-Ardila
\& Mazzaley 2006, Riffel et al. 2006). This is obvious in the nuclear region
but also visible in the off-nuclear apertures (see Figures 2 to 5). Besides that, studies like
Galliano et al. (2003) already mention the importance of the contribution of dust clouds
emission to the continuum. In order to 
properly account for this component, we have included in our spectral
base 8 Planck distributions (black-body-BB), with T ranging from 
700 to 1400 K, in steps of 100 K (For more details see Riffel et al. 2009)

Extinction is modelled by STARLIGHT as due to foreground dust,
and parametrised by the V-band extinction A$_V$. Firstly
we use the Cardeli, Clayton \& Mathis (1989 - CCM) extinction law, 
but we also compared results with the Calzetti et al. (2000 - HZ5) law.

\section{Synthesis Results and Discussions}

The spectral synthesis fitting procedure is shown in 
Figures~2 - 5 for all apertures, using the CCM extinction law. For each aperture, the
top panel shows the observed and synthetic spectra normalized to unit at 1.223 
$\micron$ (in black). The red line represents the model. The bottom panel shows the 
residual spectrum. The quality of the fits
are measured by  $\chi^2_\lambda$ and adev as defined in Cid Fernandes et al. (2004) and Riffel et al. (2009). $\chi^2_\lambda$ 
is the $\chi^2$ divided by the number of $\lambda$'s 
used in the fit and adev gives the percentage mean deviation  
$|$O$_\lambda$~-~M$_\lambda$~$| /$~O$_\lambda$ over all fitted pixes, where O$_\lambda$ is the
observed spectra and M$_\lambda$ is the model spectra obtained by the synthesis.

 Apertures 07 and 08 to the south and 15 and 16 to the north were
too noisy, and for this reason they
were added to form two larger apertures in order to improve the quality of the fit.

Following Cid Fernandes et al.
(2005b), we present our results using a condensed population vector, to take
into account noise effects that dump small differences between similar spectral components.
This is obtained by binning the population vector x into young (x$_Y$: t $\leq$ 5
$\times$ 10$^7$ yr), intermediate-age (x$_I$: 1 $\times$ 10$^8$ $\leq$ t $\leq$ 2 $\times$
10$^9$ yr) and old (x$_O$: t $>$ 2 $\times$ 10$^9$ yr) components, using the
flux distributions. The condensed vectors are plotted on the top panel of Figure 6 for each aperture, as
a function of distance to the centre. Negative distances represent the north direction
and positive distances represent the south. Following Riffel et al. (2009)
we have also binned the 
black body components into two components: cool (BB$_c$, T $\leq$ 1000 K)
and hot (BB$_h$, T $\geq$ 1100 K). 
These components are defined based on the sublimation
temperatures of silicate ($\sim$ 1000 K) and graphite ($\sim$ 1200 K) grains 
(Barvainis 1987, Granato \& Danese 1994). Condensed black body components
for each aperture are also plotted in Figure 6 as a thick red line (BB$_h$)
and dotted magenta (BB$_c$). The contribution of the power-law is also
shown (black line).

\begin{figure*} 
\includegraphics  [width=180mm]{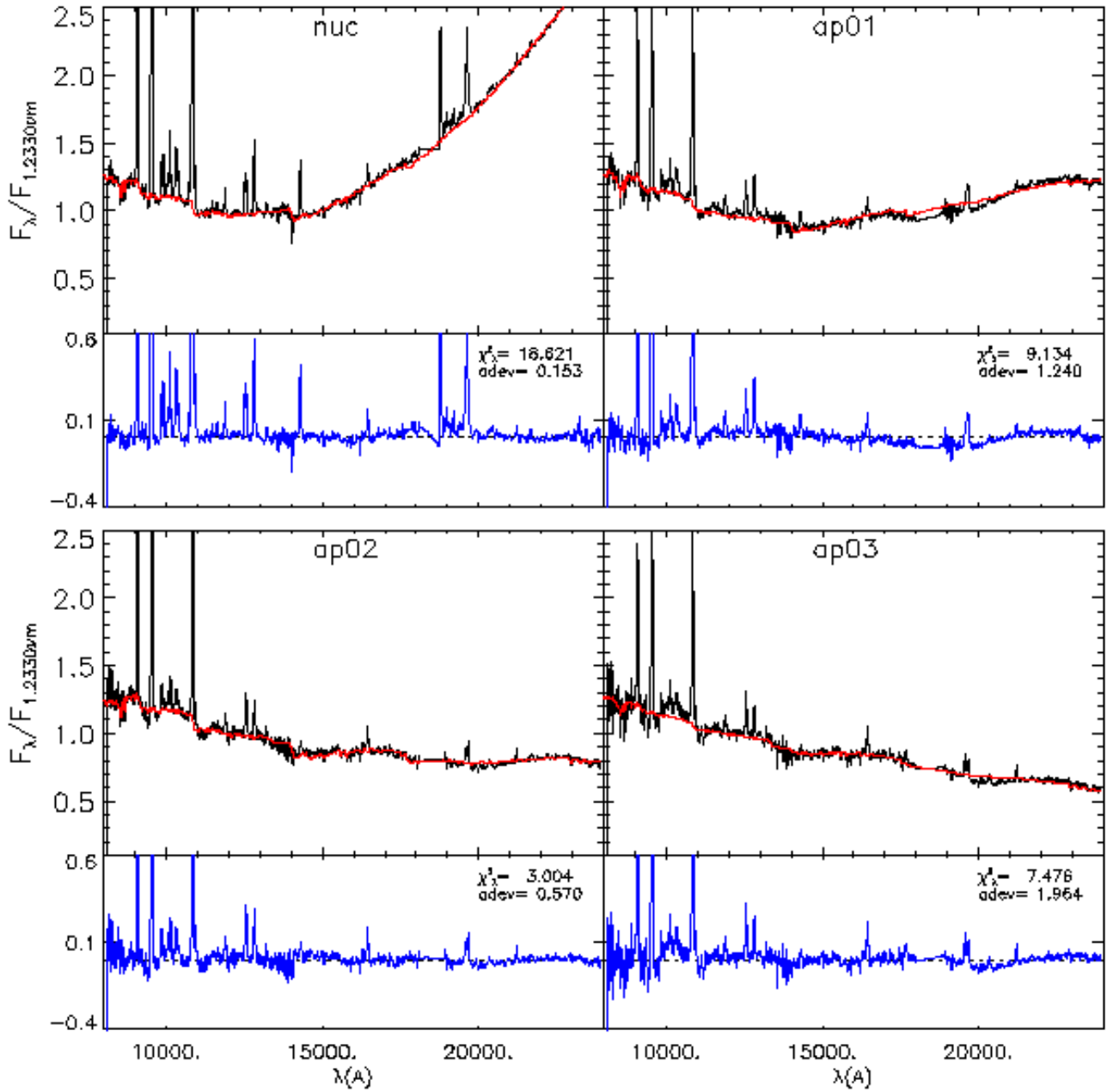}
\caption{Synthesis results for each aperture of NGC 1068.}
\end{figure*}

\begin{figure*} 
\includegraphics  [width=180mm]{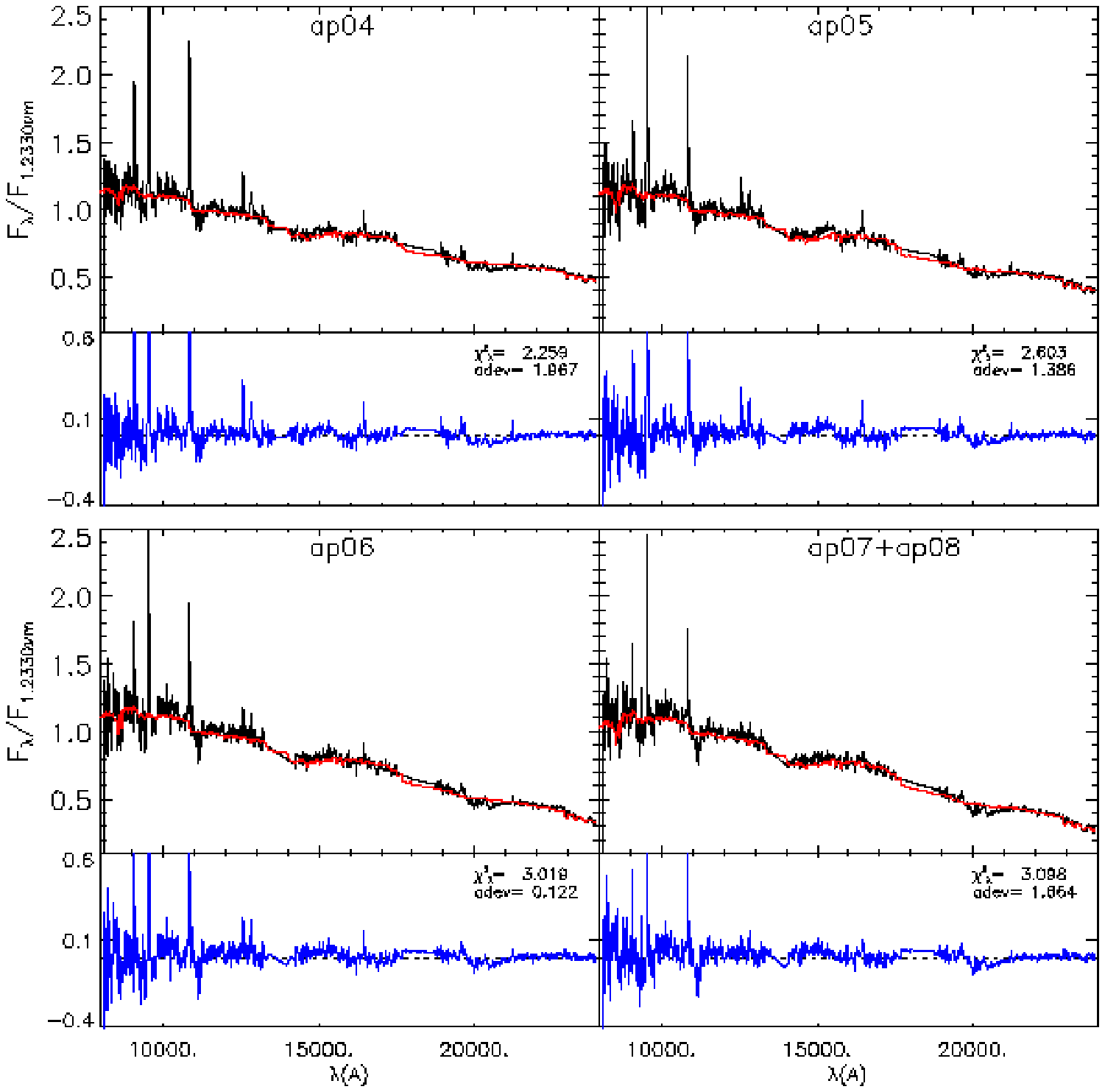}
\caption{Synthesis results for each aperture of NGC 1068.}
\end{figure*}

\begin{figure*} 
\includegraphics  [width=180mm]{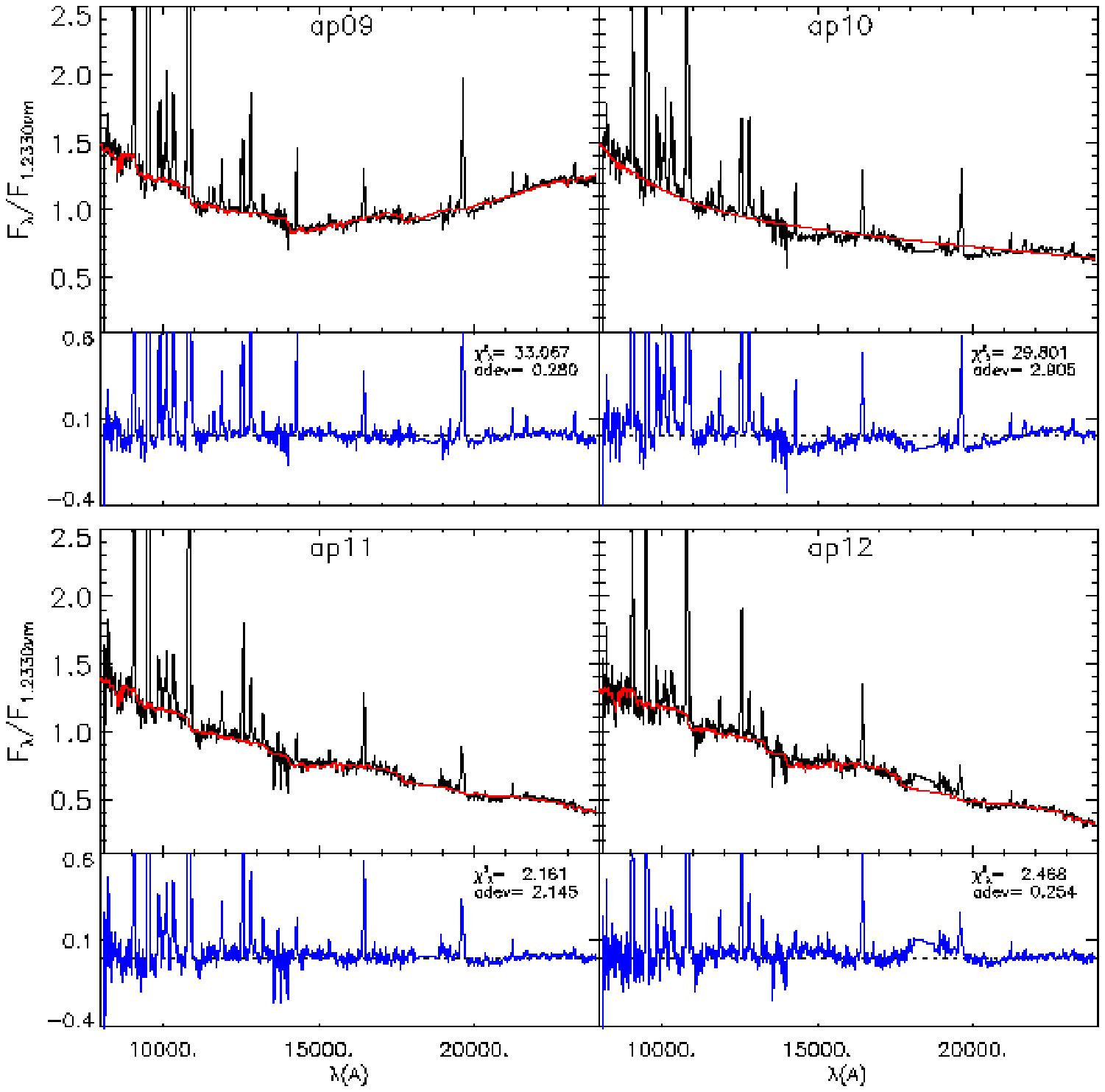}
\caption{Synthesis results for each aperture of NGC 1068.}
\end{figure*}

\begin{figure*} 
\includegraphics  [width=180mm]{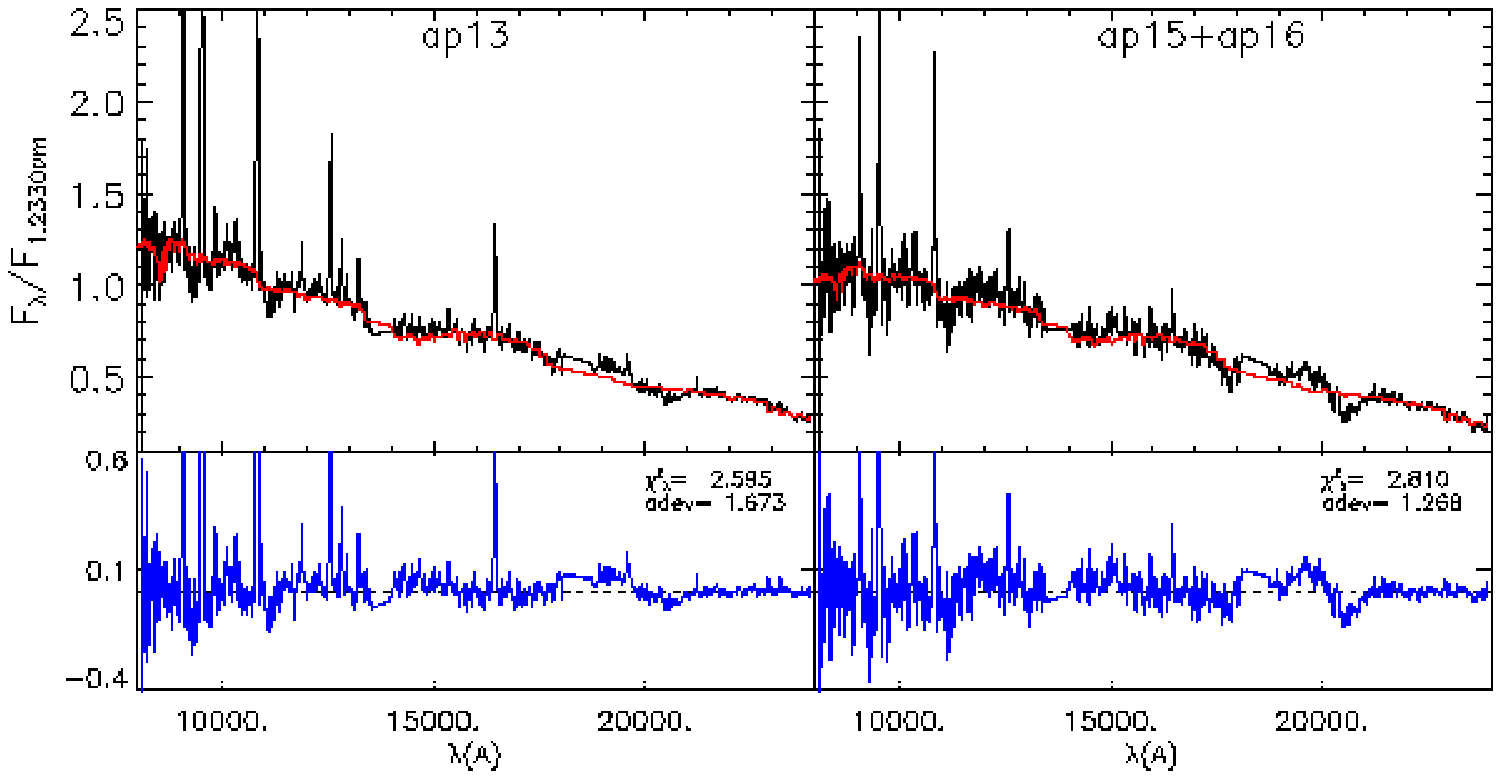}
\caption{Synthesis results for each aperture of NGC 1068.}
\end{figure*}

The spectral synthesis shows that the overall contribution of the 
stellar population to the spectra is important and in many regions 
it contributes with more than 50$\%$ of the light. The intermediate 
age population seems to be specially important. 
Indeed, one has to keep in mind that this wavelength range is specially 
sensible to this population. Thatte et al. (1997), for instance, 
found a $\sim$50~pc stellar core centred on the K-band emission peak. 
With our data we are unable to reach that spatial scale
but we could find indirect evidence confirming this result as there 
is a clear nuclear contribution of intermediate stellar population in a 
radius around 100~pc from the nucleus. Since we are integrating along
the line of sight, this population might be spreaded over a much
larger region than 100~pc. Note that the fact that
NGC~1068 is face-on minimizes this effect. Similar results are found
by Davies et al. (2007). They analysed NGC~1068, among other Seyfert galaxies, using
NIR adaptive optics integral field spectrograph SINFONI, and
found that stellar light profiles typically have size scales of a few
tens of parsecs, where there appear to have been recent, 
but no longer active starbursts. Results of Riffel et al. (2009) agree
with this scenario: Seyferts do contain a substantial fraction of intermediate-age
stellar populations. 

In the optical region, Cid Fernandes et al. (2004) applied
stellar population synthesis to NGC 1068.
They found the following values for the contribution of each
component to the nuclear spectrum: FC=23$\%$, x$_Y$= 24$\%$, x$_I$= 0$\%$ and
x$_O$= 54$\%$.
It can be seen that FC agrees with our results although they found a much higher 
contribution of the young stellar population than ours and no
intermediate age population. The reasons for this apparent contradiction
might be indeed due to the fact that the optical region is more sensible to young 
stellar populations while the NIR is better to probe intermediate age populations.
We find average ages in this region from
1 to 19 $\times$ 10$^8$~yr, consistent with the ages Cid Fernandes et al. (2004)
obtained for the stellar core
(between 5 and 16 $\times$ 10$^8$ yr).

Perhaps our most interesting result is the small young stellar population contribution
found by our stellar synthesis around $\sim$~100 pc south of the nucleus. This result, however, needs to be
taken with caution because it is
a common problem in the study of the stellar populations of active
galaxies that a reddened young starburst (age $\leq$ 5~Myr) is
indistinguishable from an AGN-type continuum (e.g. Cid Fernandes \&
Terlevich 1995, Storchi Bergman et al. 2000). In order to test this
possibility, we plot in the middle panel of 
Figure~6 the A$_V$ extinction found by the synthesis for each
aperture. It can be seen that the young population is associated with
higher extinction. It means that based only on the synthesis 
results we cannot state if the young stellar component is indeed present. However, 
other results mentioned along this section (hot dust and H$_2$ distribution) reinforce that it is indeed a real 
detection.

\begin{figure}
\includegraphics  [width=88mm]{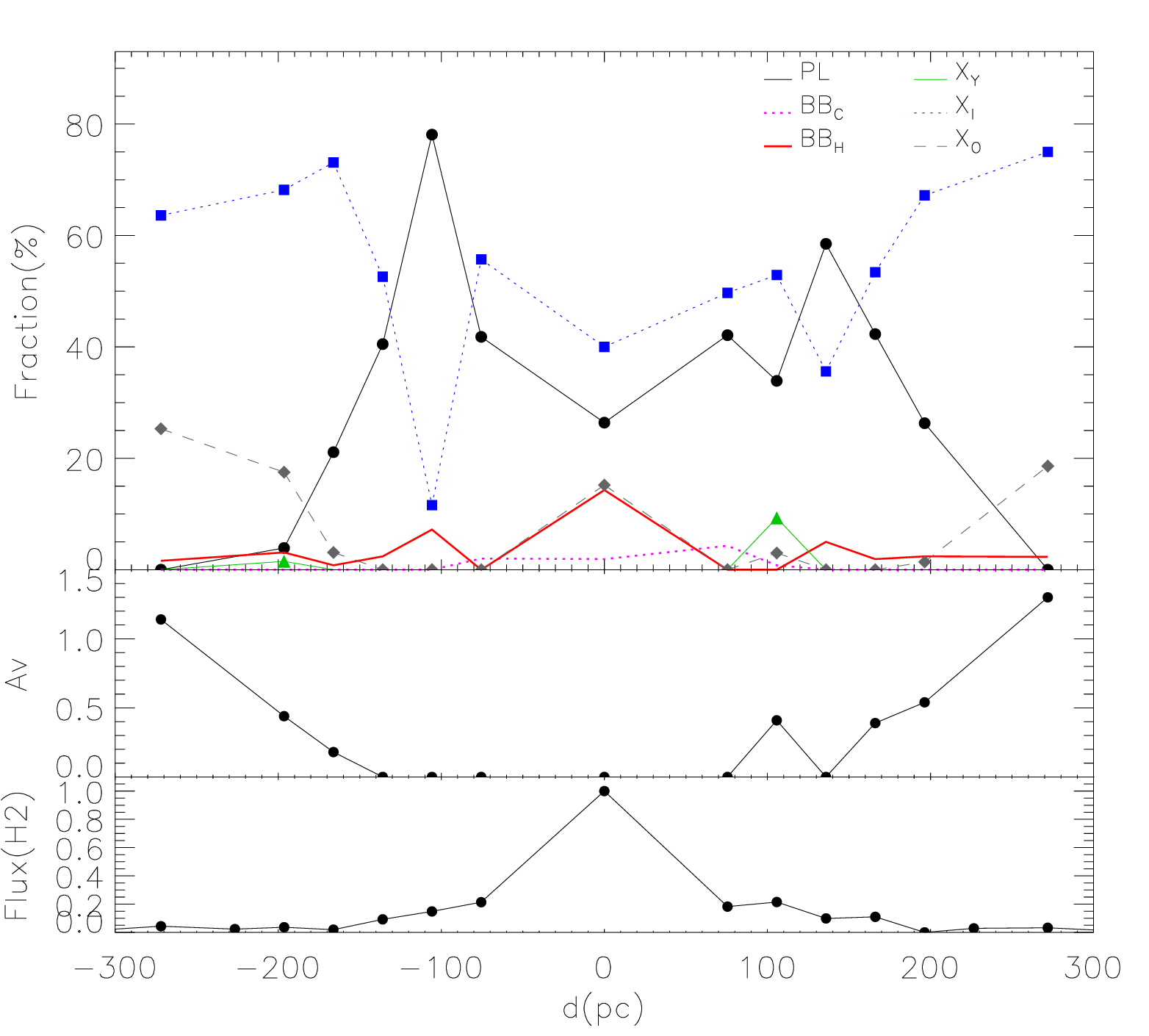}
\caption{Synthesis results as a function of distance to the nucleus.
Negative distances represent the north direction and positive distances the south.
The top panel shows the flux fraction of stellar population: x$_Y$ (young component), 
x$_I$ (intermediate-age component) and x$_O$ (old component), the flux fraction of hot dust
(BB$\_$H), cold dust (BB$\_$C) and the power law component (PL). The middle panel shows
the A$_V$ extinction obtained by the synthesis for each aperture. The bottom panel shows 
the normalized H$_2$$\lambda$2.12 $\micron$ flux obtained by Martins et al. (2010). For these
results the extinction law used was from Cardeli et al. (1989)}
\end{figure}

One way to characterize the stellar population  content of a galaxy
by a single parameter is the mean age, defined in two ways: the first 
is weighted by light fraction,
and the second is weighted by the stellar mass. Both definitions
are limited by the age range used in our element base (FC
and BB components are excluded from the sum). The mean stellar 
ages derived with both definitions for each aperture are presented
on the top panel of Figure 7. 

The light-weighted mean age of the apertures is biased to an intermediate/old
age stellar population, while for the mass-weighted the old population
clearly dominates. As stated by Cid Fernandes et. al. (2005) and Riffel et al. (2009), 
the mass-weighted mean age is a more physical parameter, but it has a much less direct 
relation with the observables. Although the mass fraction of a young
stellar population might be small, it is much more luminous. 
Thus their contribution to the luminosity is much higher. 

A secondary parameter to describe the stellar population is the metallicity,
also defined as light- and mass- weighted mean metallicity.
Both definitions are bounded by the range of $Z$ used in the base. Results
for the metallicity are also presented in the bottom
panel of Figure 7.  These results point out to a mean value around solar for both
definitions, but the light-weighted average gives higher values for the
mean metallicity than the mass-weighted. Again, the light-weighted
values are more sensitive to the younger components, while the
mass-weighted results are more sensitive to the older components.
This result is consistent with a galaxy chemical
enrichment scenario, in which the young population is enriched by the evolution
of the early massive stars.

\begin{figure} 
\includegraphics  [width=88mm]{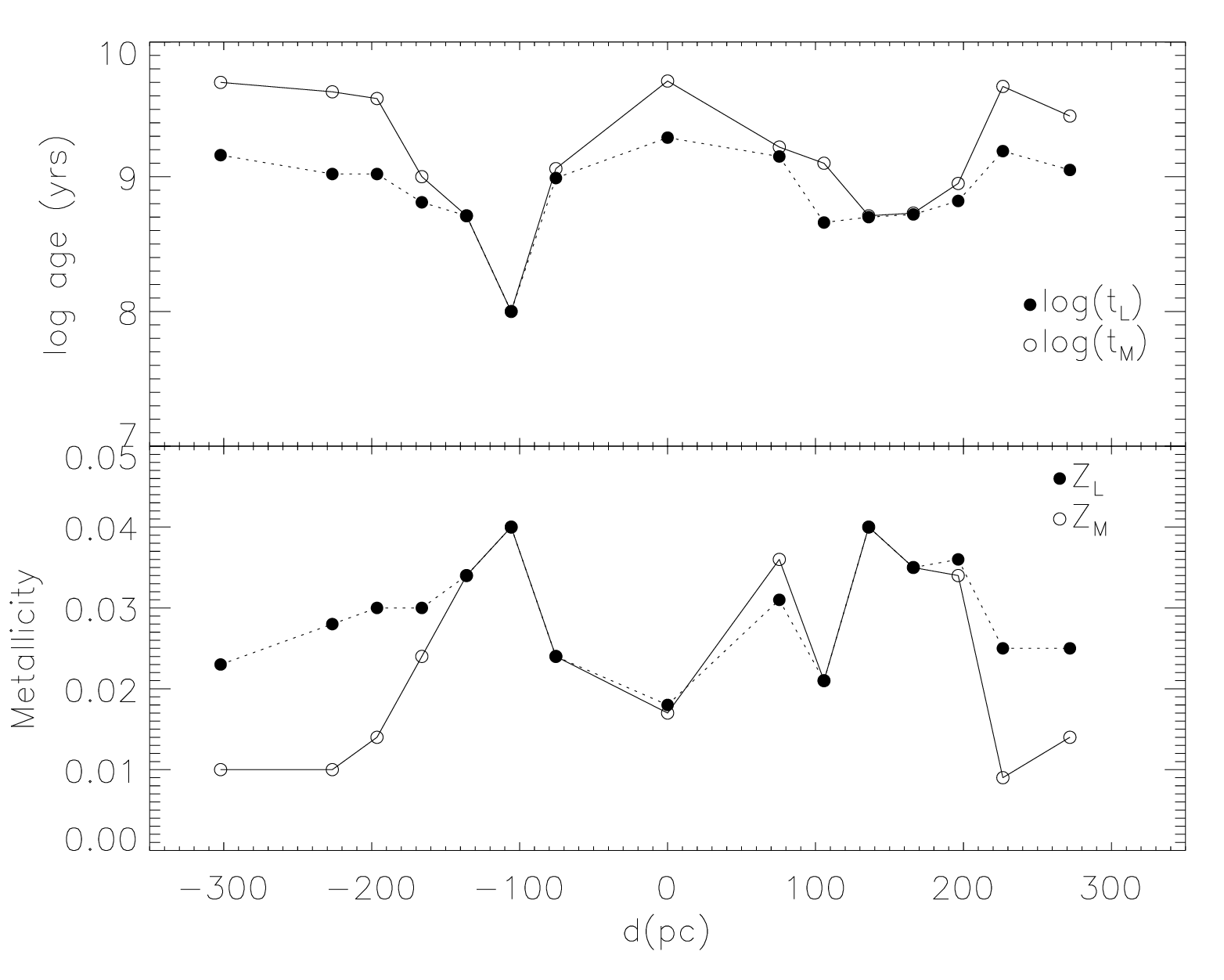}
\caption{Results for mean age and metallicity as a function of distance to the centre.
Top panel shows the light- (filled) and mass-weighted (open) mean ages. Bottom panel
shows the light- (filled) and mass-weighted (open) mean metallicities. Negative distances
represent the north direction and positive distances represent the south.}
\end{figure}

The power law component is important in the nucleus, as expected, contributing
with about 25$\%$ of the light. Cid Fernandes \& Terlevich (1995)
predicted that a broad component in H$\beta$ becomes distinguishable 
whenever the scattered FC contributes with $\geq$ 20$\%$
to the optical continuum light. Broad permitted lines are indeed observed
in the spectrum measured in polarized light, confirming that this
galaxy harbours a Seyfert 1 nucleus (Antonucci \& Miller 1985; Antonucci, Hurt, \& Miller
1994; Miller, Goodrich, \& Mathews 1991; Inglis et al. 1995; Alexander, Ruiz \& Hough 1999).
Interestingly though, this is not the region where the power-law dominates the 
spectra. There are peaks around 100 - 150 pc from  each side of the centre,
where the power-law is responsible for more than 50$\%$ of the light.
These peaks are probably 
related to shocked region. In Figure 1, these regions correspond to apertures 
03 and 10. Aperture 03, for example, is coincident with a region where the jet seems to be
changing directions, probably due to the interaction with the ISM.
If these are the
regions where the jet encounters dense clouds, shocks might have 
an important contribution to the continuum. Another possibility
is that these peaks are associated with reflected light from the BLR, due to
the higher concentration of dust. 
 In Paper I, Martins et al. 
show that many emission lines have double peaked profiles, and the maximum
of the second components are also in this region. As mentioned
above, it is important to keep in mind that the results for
the power law component might have uncertainties due to the fact that
it cannot be distinguished from a young stellar component with
dust extinction.

Hot Dust also has a significant contribution to the continuum. It
has a peak in the nuclear region, which, as reported by
Thatte et al. (1997), is a strong evidence of a dense circumnuclear
torus.
It also has peaks in the regions where the power law has a maximum, around
100 pc from the center on both sides. This is again another indication 
that these are regions where the jet finds dense
dusty/molecular clouds.  
As a further indication of this hypothesis, we  plot in bottom  panel of
Figure~6
the intensity of the H$_2$$\lambda$2.12 $\micron$ emission line, measured in Paper~I. 
This line probes the conditions of the ``warm'' molecular gas, 
and can be originated from photodissociated regions (PDR), emission due to shocks
or X-ray heated gas. 
It is interesting to notice that this emission line shows extended emission around the same region where the hot dust and 
the young stellar population are found.

Although only a small flux fraction was detected, cold dust also has an interesting
distribution. It appears to be concentrated on the southern spectra. It was 
already observed that [OIII] emission is very faint in the southwest radio cone
direction (Unger et al. 1992) and very little UV/optical polarized light
(Scarrott et al. 1991) in this region. This suggests that the southwest cone lies
behind the plane of the galaxy and is extinguished by 
the dust in the plane (e.g. Gallimore et al. 1994; Bland-Hawthorn et al. 1997), in 
agreement with the cool dust distribution we find here.

We also calculated the hot dust mass, obtaining a total value
of 3.4 $\times$ 10$^{-2}$ M$\sun$. From this value,
96$\%$ is from the central aperture, and about 3$\%$ are from
apertures 01, 02 and 09. These are apertures where the
contribution of BB$_C$ is higher. For the other apertures, 
BB$_H$ is the only one present. One have
to keep in mind that, although the luminosity fraction of
cold dust is small, their emissivity is also small, so it
requires a large mass. On the other hand, the emissivity
for very hot grains is large, so the calculated mass is 
small. It is interesting to notice, however, that the 
total mass calculated for the nuclear aperture is 
one of the highest mass values for Seyfert galaxies determined
so far (see Tab. 5 of Riffel
et al. 2009).

We tested if our results were dependent of some of the choices we made for the 
synthesis. For example, in Figure~8 we show results for the synthesis using Calzetti's
extinction law instead of CCM. Using this extinction law the contribution of the
stellar population is even clearer, and it is present in both sides of the nucleus.

\begin{figure} 
\includegraphics  [width=88mm]{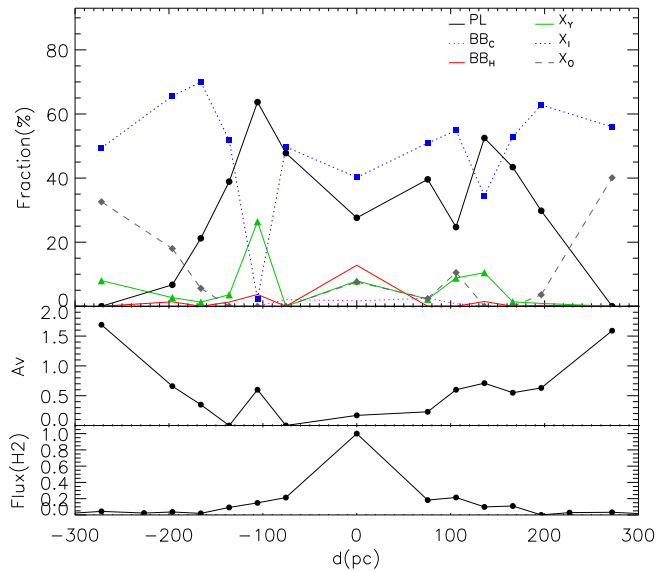}
\caption{Same as in Figure 6, but for Calzetti's extinction law.}
\end{figure}

We also tested modifying the power law index to 1.25 and 1.75 to check if the FC
contribution would change. We find that the our results are not affected by this
change.

\section{Summary and conclusions}

We investigated the NIR spectra of NGC~1068 extended 
across 15" ($\sim$ 1100 pc) in the North-South direction. 
The spectra was taken with the IRTF SpeX instrument,
obtained in the short cross-dispersed mode. 
The objective was to disentangle 
the galaxy's spectral energy distribution components along the
full wavelength coverage (0.8 $\micron$ - 2.4 $\micron$) using 
stellar population synthesis. This was done here for the first 
time for this galaxy.
We used the STARLIGHT code, which considers the
whole observed spectrum, continuum and absorption features.

We find that intermediate age stellar population contributes
significantly to the continuum, specially in the inner
200 pc. We don't have the spatial resolution to determine
the size of this nuclear stellar cluster, but results agree
with previous determinations of the stellar population of
NGC~1068 from Davies et al. (2007) and Riffel et al. (2009).

We find a small contribution of a young stellar population component about
100 pc from the nucleus. Although this result
has to be considered carefully, since the detection might
be an artefact from the method used, taken together with
our other findings this suggests that this is
the region where the jet interacts with the ISM
and might be forming stars. 

The contribution of a power-law
continuum in the centre is about 25$\%$,
which is expected from a scattered Seyfert~1 nucleus (Cid Fernandes et al. 1995). 
We also find
peaks in the contribution of the featureless continuum 
about 100 - 150 pc from the nucleus, both in the north and in the south direction.
These might be associated with the region where the
jet encounters dense clouds. This result
is further supported by
peaks of hot dust found around the same regions. 
We compared these results with the H$_2$ emission line profile
obtained from Paper I. This line also shows extended emission
around the same region where the hot dust is found.
Taking this together with the young stellar population 
detection, we believe there is strong evidence that the interaction
of the jet with the dense ISM in NGC~1068 is forming stars. Further
investigation with more spatial resolution is needed to confirm this hypothesis.

Hot dust
also has an important contribution to the nuclear region, 
which reinforces the idea of the presence of a dense, circumnuclear
torus in this galaxy. Cold dust appears mostly in the south
direction, which supports the idea that the southwest emission
is behind the plane of the galaxy, and is extinguished by
the dust in the plane. We determine the dust mass in this galaxy
to be about 3.4 $\times$ 10$^{-2}$ M$_{\sun}$, which puts NGC 1068 as one of the
Seyferts with the highest dust content measured.

\section*{Acknowledgments}
The authors thank the anonymous referee for valuable comments. This research has been partially supported by the Brazilian agency FAPESP (2007/04316-1).
R.R. thanks the Brazilian funding agency CAPES. ARA thanks to CNPq for financial support through grant 308877/2009-8.

\bsp

\label{lastpage}

\end{document}